\documentstyle[onecolumn]{mn}

\newif\ifAMStwofonts

\def\HI{\hbox{H\,{\sc i}\,}}
\def\HII{\hbox{H\,{\sc ii}\,}}

\ifoldfss
  \ifCUPmtlplainloaded \else
    \NewTextAlphabet{textbfit} {cmbxti10} {}
    \NewTextAlphabet{textbfss} {cmssbx10} {}
    \NewMathAlphabet{mathbfit} {cmbxti10} {} 
    \NewMathAlphabet{mathbfss} {cmssbx10} {} 
  \fi
  \ifAMStwofonts
    \ifCUPmtlplainloaded \else
      \NewSymbolFont{upmath} {eurm10}
      \NewSymbolFont{AMSa} {msam10}
      \NewMathSymbol{\upi}     {0}{upmath}{19}
      \NewMathSymbol{\umu}     {0}{upmath}{16}
      \NewMathSymbol{\upartial}{0}{upmath}{40}
      \NewMathSymbol{\leqslant}{3}{AMSa}{36}
      \NewMathSymbol{\geqslant}{3}{AMSa}{3E}

       \let\le=\leqslant
       \let\ge=\geqslant
    \fi
  \fi
\fi 

\ifnfssone
  \newmathalphabet{\mathit}
  \addtoversion{normal}{\mathit}{cmr}{m}{it}
  \addtoversion{bold}{\mathit}{cmr}{bx}{it}
  \newmathalphabet{\mathbfit} 
  \addtoversion{normal}{\mathbfit}{cmr}{bx}{it}
  \addtoversion{bold}{\mathbfit}{cmr}{bx}{it}
  \newmathalphabet{\mathbfss} 
  \addtoversion{normal}{\mathbfss}{cmss}{bx}{n}
  \addtoversion{bold}{\mathbfss}{cmss}{bx}{n}
  \ifAMStwofonts
    \ifCUPmtlplainloaded \else
      \UseAMStwoboldmath
      \makeatletter
      \new@mathgroup\upmath@group
      \define@mathgroup\mv@normal\upmath@group{eur}{m}{n}
      \define@mathgroup\mv@bold\upmath@group{eur}{b}{n}
      \edef\UPM{\hexnumber\upmath@group}
      \new@mathgroup\amsa@group
      \define@mathgroup\mv@normal\amsa@group{msa}{m}{n}
      \define@mathgroup\mv@bold\amsa@group{msa}{m}{n}
      \edef\AMSa{\hexnumber\amsa@group}
      \makeatother
      \mathchardef\upi="0\UPM19
      \mathchardef\umu="0\UPM16
      \mathchardef\upartial="0\UPM40
      \mathchardef\leqslant="3\AMSa36
      \mathchardef\geqslant="3\AMSa3E

       \let\le=\leqslant
       \let\ge=\geqslant
    \fi
  \fi
\fi 

\ifnfsstwo
  \DeclareMathAlphabet{\mathbfit}{OT1}{cmr}{bx}{it}
  \SetMathAlphabet\mathbfit{bold}{OT1}{cmr}{bx}{it}
  \DeclareMathAlphabet{\mathbfss}{OT1}{cmss}{bx}{n}
  \SetMathAlphabet\mathbfss{bold}{OT1}{cmss}{bx}{n}
  \ifAMStwofonts
    \ifCUPmtlplainloaded \else
      \DeclareSymbolFont{UPM}{U}{eur}{m}{n}
      \SetSymbolFont{UPM}{bold}{U}{eur}{b}{n}
      \DeclareSymbolFont{AMSa}{U}{msa}{m}{n}
      \DeclareMathSymbol{\upi}{0}{UPM}{"19}
      \DeclareMathSymbol{\umu}{0}{UPM}{"16}
      \DeclareMathSymbol{\upartial}{0}{UPM}{"40}
      \DeclareMathSymbol{\leqslant}{3}{AMSa}{"36}
      \DeclareMathSymbol{\geqslant}{3}{AMSa}{"3E}

       \let\le=\leqslant
       \let\ge=\geqslant
    \fi
  \fi
\fi 

\ifCUPmtlplainloaded \else
  \ifAMStwofonts \else 
    \def\upi{\pi}
    \def\umu{\mu}
    \def\upartial{\partial}
  \fi
\fi
\def\simlt{\lower.5ex\hbox{$\; \buildrel < \over \sim \;$}}
\def\simgt{\lower.5ex\hbox{$\; \buildrel > \over \sim \;$}}

\title{The Extended Rotation Curve and the Dark Matter Halo of M33}

\author[E. Corbelli and P. Salucci ]
{Edvige Corbelli$^1$, Paolo Salucci$^2$
\\
 $^1$ Osservatorio Astrofisico di Arcetri, Largo E. Fermi,5
I-50125 Firenze, Italy\\
$^2$ International School for Advanced Studies, SISSA,
Via Beirut 2-4, I-34013 Trieste, Italy\\ 
\      
         }

\date{Received .... ; accepted .....}

\begin{document}

\maketitle  

\label{firstpage}

\begin{abstract}

We present the 21-cm rotation curve of the nearby galaxy M33 out to
a galactocentric distance of 16 kpc (13 disk scale-lengths). 
The rotation curve keeps rising out to the last measured point and implies
a dark halo mass $\simgt 5\times 10^{10}$ M$_\odot$. 
The stellar and gaseous disks provide virtually equal contributions 
to the galaxy gravitational potential at large  
galactocentric radii but no obvious correlation is found 
between the radial distribution of dark matter and the distribution of 
stars or gas. 

Results of the best fit to the mass distribution in M33 picture 
a dark halo which controls the gravitational potential from 3 kpc outward,   
with a matter density which decreases radially as $R^{-1.3}$. The
density profile is consistent with the theoretical predictions for 
structure formation in hierarchical clustering cold dark matter models 
and favors lower mass concentrations than those expected in the  
standard cosmogony.

\end{abstract}


\section{Introduction}

The ubiquitous presence of dark matter in spiral galaxies is well established 
(Rubin, Ford $\&$ Thonnard 1980, Bosma 1980) as is the increase of the dark  
matter fraction with decreasing luminosity (Persic $\&$ Salucci 1988, 1990).  
However, crucial issues at the heart of galaxy formation theories  
are still open, such as the actual density profile of the dark halo from 
the center of the galaxy out to its virial radius, or the nature of the dark 
matter. In particular, the Cold Dark Matter theory predicts a well-defined 
radial density profile for the collision less particles which make up the dark 
halo (e.g. Navarro, Frenk $\&$ White 1997); 
this should be immediately compared with  
that derived from observations of nearby galaxies. On the other hand, a strong 
link between the gas and the dark matter distribution could expose the baryonic
nature of the latter.

The main problems in tackling these issues originate from ambiguities in the 
mass model such as the interplay between the stellar disk and the dark halo 
at small radii, and  the presence of a gaseous disk  of poorly known 
orientation, size, and  surface density at large radii. It is for this 
reason that examining the nearby dwarf spiral M33 has decisive
advantages:  $(a)$ this galaxy is a primary distance calibrator and 
its distance is well determined and independent of $H_0$ ($D= 0.75\pm 0.1$
 Mpc, e.g. Freedman, Wilson $\&$ Madore 1991 and references therein).
$(b)$ Being one of the nearest galaxies at low declination it
can be observed with the Arecibo single-dish telescope that provides good 
sensitivity and a reasonable angular resolution (3.8 arcmin). 
This allows the sampling of regions very far from the center where the
\HI column density has declined to $(1-2)\times 10^{19}$ cm$^{-2}$. 
Corbelli $\&$ Schneider (1997, hereafter CS) have carefully investigated 
the geometry of the warped gaseous disk from the \HI fluxes measured. 
$(c)$ M33 is a normal low-luminosity dark matter-dominated spiral (e.g.  
Persic, Salucci $\&$ Stel 1996, hereafter PSS) in which it is relatively 
easy to disentangle the dark and luminous mass components.

The aim of this paper is to derive the rotation curve of M33 from the \HI
data and to determine the characteristics of the dark matter distribution.  
In Section 2 we investigate the surface density profile of stars and gas and 
in Section 3 we derive the rotation curve and analyze the possible
signatures of interaction with its neighbor M31. The main structural 
properties of the dark halo are presented in a model-independent way in
Section 4, while in Section 5 we derive a more detailed mass model
and check its consistency with galaxy formation calculations.

\section{The density distribution of the identified baryonic matter}

In M33 there are four ``luminous'' components that could contribute to 
the gravitational potential:

\noindent           
{\it (a) Stellar disk}
                         
A thin disk is the main stellar component since the central bulge  
is very small and can be completely neglected. For the stellar
disk we use the exponential scale-length,
$R_d$, as measured in the K-band. This more closely reflects the underlying 
stellar mass distribution and is less affected by extinction: 
$R_d \simeq 5.8$ arcmin $\simeq 1.2 \pm 0.2 $ kpc (Regan $\&$ Vogel 1994).
As in the rest of this paper we have assumed the following conversion: 
5 arcmin = 1 kpc corresponding to a distance $D\simeq 0.7$ Mpc. 
M33 is an extremely blue galaxy having
$(B-V)^0_T=0.46$ (de Vaucouleurs at al. 1991). The total blue luminosity
in units of blue solar luminosity is $L_B=4.2\times 10^9\ {\hbox{L}}
_\odot$ (Sandage $\&$ Tamman 1981).  

\noindent
{\it (b) Atomic gas}

Most of the gaseous mass in M33 is in the form of neutral atomic hydrogen.  The
high sensitivity observations of M33 made with the Arecibo 305-m radiotelescope
have allowed  CS  to draw a detailed
map of the spatial extent of the neutral gaseous component down to a limiting
column density $N_{\HI}\simeq 1-2\times 10^{19}$ cm$^{-2}$.  The results of the
detailed survey of the outer disk and of the tilted-ring model fitted to the
21-cm line data have shown a radial extent of the \HI which is more than
13 times the exponential scale-length of the stellar disk.  The
outer disk of neutral hydrogen in M33 is warped and oriented at $\sim 30\deg$
with respect to the inner disk and the global \HI profile of the outer 
disk appears symmetric on the low and high velocity wings. The \HI surface
density profile and the warp model are presented in the next
Section. The total \HI mass is estimated to be
1.8$\times 10^9$ M$_\odot$ (assuming D=0.7 Mpc)  25 per cent of which 
resides in the outer disk.  

\noindent
{\it (c) Warm Ionized gas}

If the background ionizing radiation accounts for the sharp \HI fall off seen
in the outer disk around $3\times 10^{19}$ cm$^{-2}$ (Corbelli $\&$ Salpeter
1993),  
a similar amount of ionized gas is expected to lie above and below the whole
\HI disk since this is exposed to the same background radiation field.
If we take a column density of ionized gas equal to
$3\times 10^{19}$ cm$^{-2}$  ($\sigma_{\HII}$=0.26 M$_\odot$ 
pc$^{-2}$) throughout the \HI disk we have an \HII mass of 1.9
$10^8$ M$_\odot$. This is only a rough estimate since we neglected
radiation and gravity from the stellar disk, but it 
agrees with the surface density of the
ionized gas detected in the outer parts of  
NGC 253 (Bland-Hawthorn, Freeman $\&$ Quinn 1997) and above the disk of 
NGC 891 (Dettmar 1992). Since the detailed radial distribution of 
the warm ionized gas is unknown and its estimated mass is only 10 per 
cent of the \HI mass 
we shall neglect its contribution to the gravitational potential.

\noindent
{\it (d) Molecular gas}

M33 is known to be a galaxy deficient in molecular gas.
Maps of the diffuse CO component and interferometric studies show that 
molecules are not a dominant component of the global gas mass fraction,   
although individual large molecular complexes with masses of order $10^{5-6}$
M$_\odot$ are prominent in the nuclear region (e.g. Young $\&$ Scoville 1982; 
Wilson $\&$ Scoville 1989). Within the first kiloparsec the derived 
H$_2$ column density is radially constant and of the same 
order of the \HI column density, but at larger radii it drops rapidly.  
At 2.5 kpc the H$_2$ mass is $6\times 10^7$ M$_\odot$, about half of 
the \HI mass. Therefore the molecular contribution to the 
potential well is small and it is a reasonable approximation to consider 
the H$_2$ surface density scale-length of the same order of the
K-band stellar scale-length $R_d$, and the molecular gas as a small part of 
the mass which we shall find for the stellar disk.

\section {The \HI disk and the outer rotation curve }

We can derive in a self-consistent  way  the \HI surface density profile  
and the related rotation curve from the observed \HI fluxes by using a 
model which corrects for the inclination of the galaxy disk with respect
to our line of sight at different radii.
For the model we use 11 free equally-spaced rings 
and to each ring we assign a value of the surface brightness,
of the rotational velocity, of the inclination and position angle
which were allowed to vary independently. Between each of these rings  
we then considered 10 additional rings whose properties were derived
using a linear interpolation of the properties of the two nearest
free rings. The position and systemic
velocities of the ring centers were also allowed to vary in order
to correct for possible tidal perturbations from M31.  
To stabilize the free ring model we surrounded the galaxy with artificial
zero flux observation. We have removed from the
data 10 observed points which make up the region of 1 kpc in  linear
extent around RA 01$^h$ 30$^m$ DEC 29$^\circ$ 24$^\prime$. This region shows  
an anomalous velocity field relative to the surrounding areas and
furthermore lies outside the region covered by the ring model
(see Fig.6 of CS). As noted by Huchtmeier (1978) the excess of velocity
and its mass are typical of a high velocity cloud complex, possibly
associated with M33. We define the $\chi^2$ and carry out a minimization 
procedure as in CS starting from the free ring models displayed in their 
Fig. 5. The model which shows the lowest $\chi^2$ demands a variation of 
-10 km s$^{-1}$ in the systemic velocity for one of the outermost rings.
This is because in the northern part of the galaxy there is a 
high velocity cloud complex superimposed on the normal \HI disk 
(see the anomalous velocity contour at -240 km s$^{-1}$ in Fig. 3
of Corbelli, Schneider $\&$ Salpeter 1989). In order to avoid a bad 
deconvolution for the rest of data points in the ring we decided to
limit the velocity shifts of the systemic velocity to be less than 5
km s$^{-1}$.
Figure 1 shows the radial variations of the inclination and position angle 
of the resulting tilted ring model which we use for deconvolving the data.  
The center shift is negligible
for the inner rings and is about 4 arcmin for the three outermost free rings.

\begin{figure}
 
\vspace{7.5cm}
\includegraphics{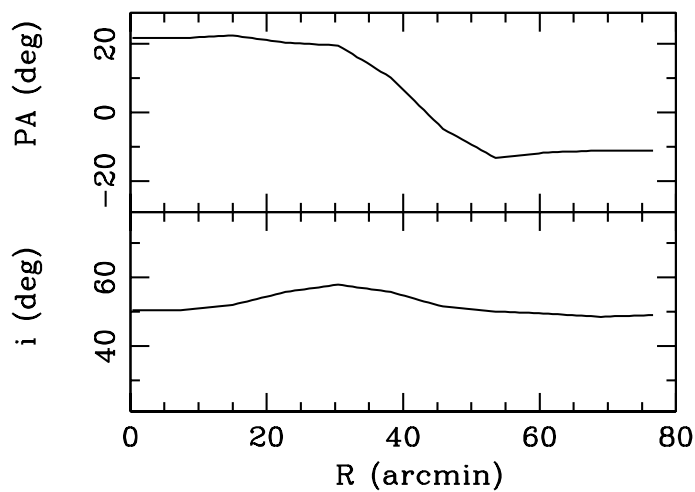}
\caption{ Inclination and position angle of the tilted ring model 
used for the deconvolution of the \HI data. }
\label{fig:fig1 }
\end{figure}

The rotational velocities are obtained by deconvolving, with the above 
geometrical model, the 320 line profiles of the emission located  at 
angles less than 45 deg with respect to the major axis and with peak fluxes
above 2000 mJy beam$^{-1}$.  The observed velocity is defined to be $V_{gau}$,
the peak of the fitted gaussian function (where double-horned
profiles were detected, $V_{gau}$ was computed as the average values of the
peaks of the two gaussians).  The choice of $V_{gau}$ minimizes the presence 
of faint secondary components and sidelobe or galactic contamination.  
The resulting rotational velocities are shown in Figure 2$(a)$. At each
radius the dispersion is low and differences between the
northern and southern velocity data sets are less marked than
what the displayed points might suggest. In fact the low velocity data  
belonging to the northern half of M33 around $R\sim 50$ arcmin come
from the region where the anomalous velocity complex is.
Similarly, the low velocity data in the southern half 
of M33 at $R\sim 60$ arcmin, lie next to the high velocity cloud
complex removed from the data. In spite of the possible 
contamination, these observations are kept in the data since they also contain  
part of the information relative to the normal outer \HI disk. This
will strengthen our conclusion about the monotonic increase
of rotation speed from the center of M33 outwards.
The straight lines in Fig.2$(a)$ are the best linear fits for $R\ge 25$ arcmin
to all data shown, to the northern data set and to the southern data set. The fits
for the two separate halves of M33  have slopes consistent, within 1-$\sigma$ errors,
with the slope relative to all data points: 0.5 $\pm$ 0.06. Similar slopes
and consistencies are found if we take out the contaminated data;
in this case no differences are found in the slopes by selecting
all data with $R\ge 25$ arcmin or data between 25 and a cutoff radius.

\begin{figure}

\vspace{13cm}
\includegraphics{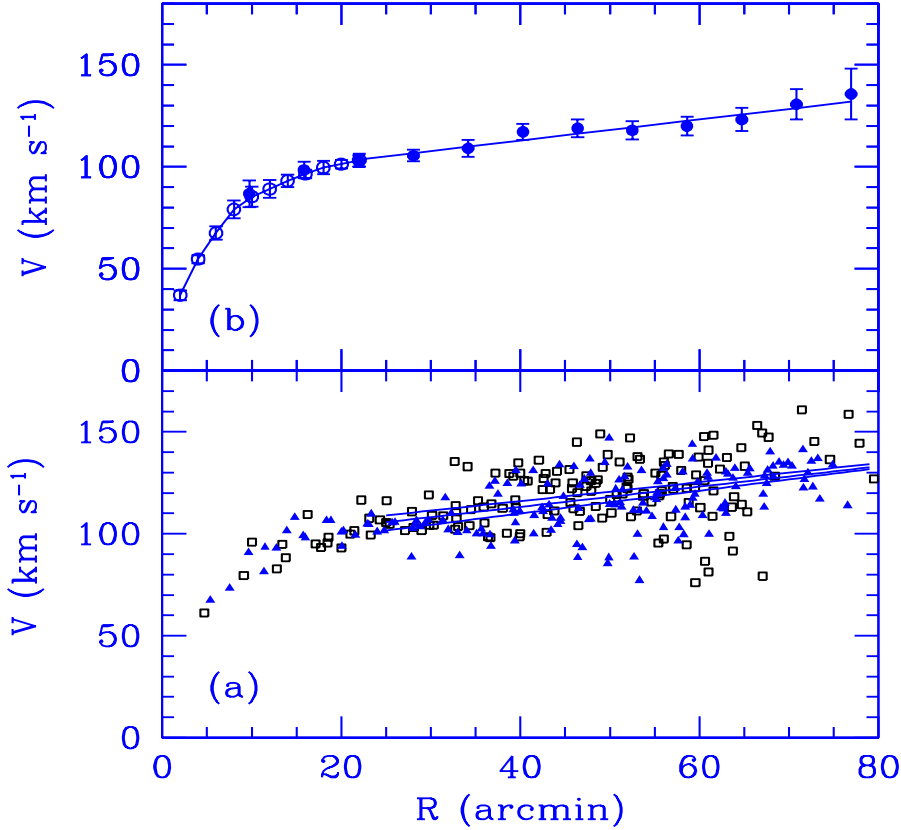}
\caption{ {\it (a)} Deconvolved velocity data. Filled triangles refer to  
the northern half of the galaxy,  open squares to  the southern half. 
Straight lines are the best linear fits for $R\ge 25$ arcmin
to all data shown, to the northern data set and to the southern data set.
{\it (b)} The M33 rotation curve: filled circles are from data displayed
in ($a$), open circles are Newton's data. The straight line is the
best linear fit to all unbinned data for $R>25$ arcmin.}
\label{fig:fig2 }
\end{figure}

The linear fit to all rotational velocities for $R\ge 25$ arcmin is also   
shown in Figure 2$(b)$ where the data from 6 to 80 arcmin have been
binned into 12 equally spaced velocity bins and displayed with
the relative $\pm$ 2-$\sigma$ Poisson errors around the mean.   
In the inner region, $R \le 20$ kpc, the present rotation curve has been
supplemented with the aperture synthesis 2 arcmin resolution rotation curve by
Newton (1980).  Figure $2(b)$ illustrates the good match 
between the two data sets. 

In conclusion, the present rotation curve of M33 extends out to 16 kpc 
at a resolution of 0.4-1.2 kpc, with a 4 per cent mean amplitude
uncertainty.  Such a rotation curve 
is virtually unprecedented for its radial extension in units of disk 
scale-lengths being the outermost data point at about 13 $R_d$. 
The data show no clear signs of kinematic disturbances from M31
in terms of north-south asymmetries or induced transient features in the 
rotation profiles from a closer encounter in the past
(Keel 1993, Barton, Bromley $\&$ Geller 1999)
but higher resolution data in the optical region are more appropriate for 
addressing this issue. As discussed by CS, M33 is now outside the Roche 
limit of M31 and we expect only weak tidal disturbances like the observed
twist of the outer disk or elliptical orbits. For the outermost envelope
of M33, at a radius of 1/12th the separation between M33 and M31, the
equipotential surface is offset toward M31 by about 4 arcmin. Despite the
uncertainties in this number, which reflect the uncertainties in the
mass distribution around M33 and M31, it is remarkable how it agrees
with what we found for the center shift of the three outermost rings 
in the tilted ring model.

The resulting radial distribution of the \HI gas is shown in Figure 3.
The filled circles are obtained by averaging the data points inside
equally spaced rings of 3.5 arcmin width.  The line is the analytic fit
used to compute the \HI gas contribution to the rotation curve.  We assume
that the gas distribution falls off exponentially beyond 80 arcmin, where 
the \HI reaches the observational sensitivity limit, and we correct for 
the primordial helium abundance by multiplying the observed neutral 
hydrogen surface density by 1.33.

\begin{figure}
 
\vspace{8cm}
\includegraphics{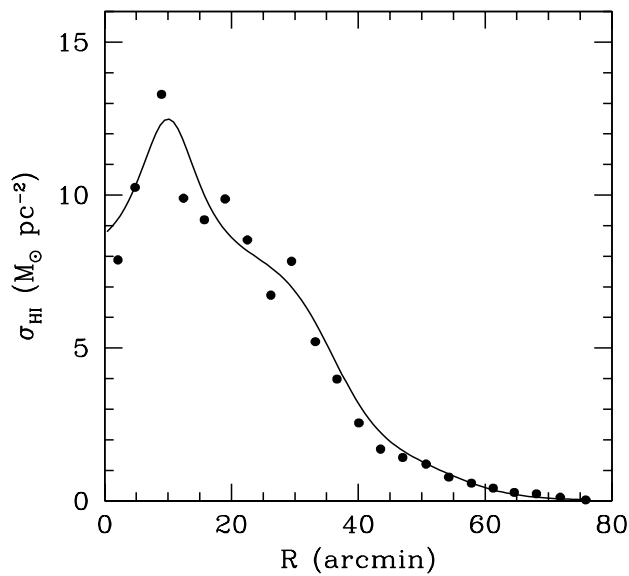}
\caption{ Averaged data (filled dots) and analytic fit to the surface
density of neutral hydrogen as a function of radius.}
\label{fig:fig3 }
\end{figure}

From the rotation curve and from the \HI+He surface density,
we can derive the classical disk stability  
parameter $Q$ (Toomre 1964, Kennicutt 1989).
Values of $Q<1$ would in principle indicate the presence of a large scale 
star formation. For a dispersion velocity of 9 km s$^{-1}$, we find 
$Q\sim 2$ from 2 to 7 kpc, and beyond 7 kpc it diverges, reaching the 
value of 6 at $R\sim 10$ kpc. In the innermost region, Wilson, Scoville $\&$
Rice (1991) have also found $Q>1$ considering the additional H$_2$ surface density.
Even using the total surface density (gas + a stellar disk with $M_d/L_B=0.8$,
see Section 5) $Q<1$ only for $R\simlt 4$ kpc. 
It is therefore questionable whether in this galaxy a global disk 
gravitational instability is triggering the star formation activity.
One possibility is that a new disk stability parameter is regulating the star 
formation in this galaxy, as suggested by  Elmegreen (1993) and by Pandey $\&$
van de Bruck (1999).

\section{Model-independent dark matter properties }

The \HI+He disk contribution to $V(R)$, 
$V_g(R)$, is computed by means of the radial distributions  discussed 
in the previous Section and  by assuming a vertical thickness 
of 0.5 kpc (result are unchanged by taking an infinitesimally thin disk). 
The gas contribution increases slowly from the galaxy center out to $\simeq 8
{\hbox { kpc}}$ where it peaks with $V_g \sim 40$ km s$^{-1}$.  At larger radii, 
$V_g$ decreases and reaches $V_g \simeq 30$ km s$^{-1}$ at the farthest radius with 
the available \HI data. Figure 4 shows,
in terms of the parameter $\beta_g(R)\equiv V^2_g(R)/V^2(R)$,
that $V^2_g (R) << V^2(R)$ and that $\beta_g(R)$ is far from being constant.  It 
is then evident that in this galaxy the gas distribution does not trace the mass
distribution.

\begin{figure}
 
\vspace{6cm}
\includegraphics{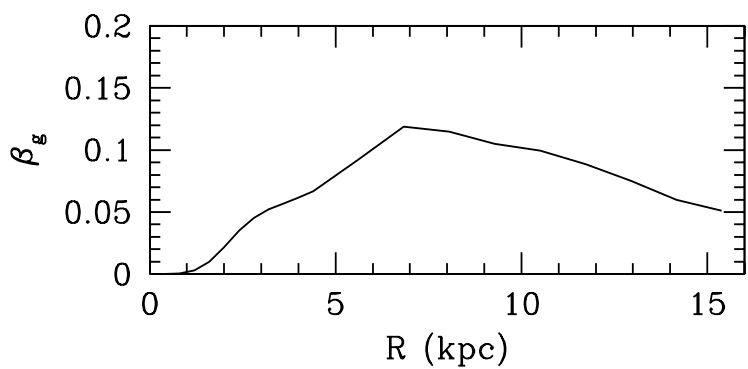}
\caption{ Fractional contribution of the gaseous disk to $V$ }
\label{fig:fig4 }
\end{figure}

We shall now work out the main model-independent properties of
the ``luminous'' matter contribution to the rotation curve.  
The blue color of this galaxy, $(B-V)^0_T=0.46$, strongly suggests  that $0.5< 
M_d/L_B<1.5$ (e.g. Tinsley 1981) with $L_B$ well determined since the distance to
this galaxy is known. The stellar disk contribution $V_d(R)$ to the circular 
velocity can be written as:

\begin{equation}
V_d^2(x)=G {M_d\over {2R_d }} x^2 B(x/2) \qquad {\hbox{with }}B(x)\equiv
I_0(x)K_0(x)-I_1(x)K_1(x)
\end{equation}

\noindent 
where $x\equiv R/R_d$, $I_0,K_0,I_1,K_1 $ are the modified Bessel functions
and $M_d$ is the total disk mass due to stars and molecular gas.
The gas contribution is an important fraction
of $V_{lum}\equiv\sqrt{V^2_d(R)+V^2_g (R)}$,
and the radial distribution of the \HI mass is known 
with uncertainties of order 20 per cent (see CS). Therefore we can 
estimate $V_{lum}(R )$ with the reasonable uncertainty of
30-50 per cent.  For virtually any other spiral galaxy
uncertainties in the stellar and
gas mass are much larger. Estimated upper and lower limits to
$V_{lum}(R)$ are shown in Figure 5. The upper curve is
derived by assuming $M_d/L_B = 1.5$ with the neutral gas surface density
increased by 20 per cent (relative to Figure 3).  For
the lower curve $M_d/L_B = 0.5$ and the neutral gas surface density is decreased
by 20 per cent.  In the region 3 kpc $ < R < $ 8 kpc, $V_{lum}(R)$ decreases 
at a rate only weakly dependent on the exact value of $M_d/L_B$ while, for
$R >8$ kpc and independently of the value of $M_d/L_B$, $V_{lum}$ enters into 
the Keplerian regime:  $V_{lum}(R) \propto R^{-1/2} $.  On the other hand, the
observed rotation curve beyond 3 kpc is $V(R) \gg V_{lum}$ (Fig. 2), 
and shows different and very well distinct features:  $V(R)$ increases 
by $\sim 20$ km s$^{-1}$ across the
region $3 {\hbox { kpc}} <R< 8 $ kpc and by $\sim 10$ km s$^{-1}$
from 8 to 16 kpc.
From these considerations it emerges that the M33 mass distribution is largely 
dominated by a non-luminous component down to $R \sim 3 {\hbox { kpc}}$.

\begin{figure}
 
\vspace{8cm}
\includegraphics{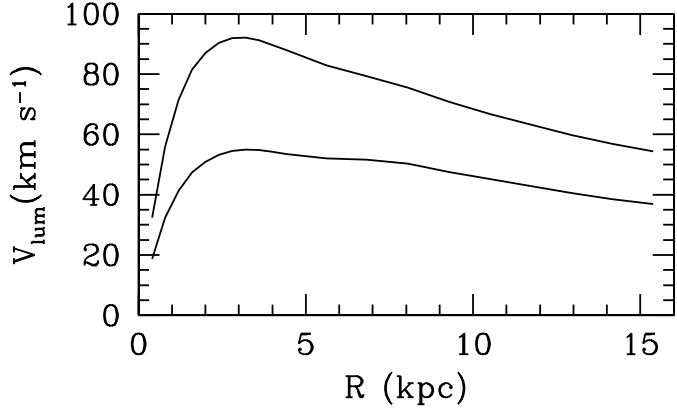}
\caption{Maximum and minimum luminous (stars + gas) contribution to $V(R)$ }
\label{fig:fig5 }
\end{figure}
 
We can estimate the properties of the dark halo in the outer disk 
using the curves of Figure 5 as upper and lower
limits for the ``luminous'' matter contribution   
to the gravitational potential. If the 
effective decline of the dark matter density in the halo is 
$\rho_h(R)\propto R^{\alpha}$, $\alpha$ can be written 
as a function of $\beta_{lum}(R)\equiv V^2_{lum}(R)/V^2(R)$,
of $d{\hbox{log}} V/d {\hbox{log}}R$, and of $d{\hbox{log}} V_{lum}
/d {\hbox{log}}R$ using the condition for centrifugal equilibrium and 
its first moment (Persic $\&$ Salucci 1990). In the outermost regions of
M33 we find $\alpha=1.3 \pm 0.15$ and therefore we can exclude 
that locally the dark matter density declines as fast as $R^{-2}$. 

\section{The mass model}

Using a large sample of rotation curves of galaxies derived for 
$R\simlt 2 R_{opt}$,
PSS presented evidence for a universal non-singular isothermal halo,
with a core radius which increases with galaxy total luminosity and
a central density which scales inversely with luminosity.
The rotation curve for M33 
is similar to that of galaxies of the same luminosity 
for $R\simlt 2 R_{opt}$ (e.g. PSS), but it gives the opportunity to
test the predictions for a isothermal halo since it extends out
to $R\simeq 4 R_{opt}$. The PSS isothermal halo model predicts a flat 
rotation curve for $R\gg R_{opt}$ while we have seen that
the rotation curve for M33 rises steadily. Non-singular isothermal spheres
have other difficulties between which the poor fits 
they provide to the structure of cold dark matter halos formed in a 
hierarchically clustering universe (e.g. Navarro, Frenk $\&$ 
White 1997). For these reasons it is necessary to assume
a more general form of dark halo density which includes also the possibility
of a non-singular isothermal sphere, with a varying amplitude of the core 
radius, and models with a non-isothermal large scale behavior.  
 
Taking into account the results of the previous Sections, the mass model 
for M33 includes: a thin exponential stellar disk, a gaseous 
disk and a dark halo.
Introducing  the variable $y\equiv R/R_{opt}$, which we shall use in the
rest of this Section, from equation (1) the contribution of the stellar disk 
can be written as:

\begin{equation}
V^2_d(y)=V^2(1) \beta_d(1) {B(1.6y)\over B(1.6)} y^2 
\end{equation}

\noindent
with $\beta_d(y)\equiv V^2_d(y)/V^2(y)$. The dark  
halo contribution to the circular velocity is parameterized as:
 
\begin{equation}
  V^2_h(y)=V_h^2(1)(1+a^2) {y^2\over y^c+a^2} 
\end{equation}

\noindent 
where  $V_h^2(1) = [1-\beta_{g}(1)-\beta_d(1)]\ V^2(1)$, 
$a$ is a constant, and $c$ indicates 
the decline of the dark matter density at large radii, for $y^c\gg a^2$. 
A least square method is used to determine the three parameters $a$, 
$\beta_d(1)$, and $c$, from the data.
This procedure allows, without forcing into them, a variety of 
mass models, including the halo-dominated and the maximum-disk model, the 
no-core halo model and the non-singular isothermal halo with a flat 
rotation curve at large radii ($c=2$, $a\ne0$). It will be shown at the
end of this Section that it will also allow us to test the
density profiles derived from cosmological N-body
simulations of structure formation in a hierarchically clustering universe.

\subsection
{The best-fit mass model}

The measured variable $w^{obs}$ and the model
function $w^{mod}$ for the least square method
are defined as follows:

\begin{equation}
w^{obs}_j\equiv \sqrt{V^2(y_j)-V_g^2(y_j)\over V^2(1)}
\end{equation}

\begin{equation}
w^{mod}_j\equiv\sqrt{[1-\beta_{g}(1)-\beta_d(1)](1+a^2) {y_j^2\over y_j^c+a^2}
+\beta_d(1){B(1.6y_j)\over B(1.6)}y_j^2  }
\end{equation}

\noindent
The $\chi^2$ is given by:

\begin{equation}
\chi^2=\sum_{j=1}^{n}{(w^{obs}_j-w^{mod}_j)^2\over
\sigma_j^2}  
\end{equation}
\smallskip

\noindent
We minimize $\chi^2$ over the three parameters $a,\beta_d(1),c$
using the binned
data shown in Figure 2$(b)$ ($n=20$, Newton data for $R\le 4.5$ kpc, 
CS data at larger radii). The errors $\sigma_j$ on $w^{obs}$ are 
functions of the errors on $V(y_j)$, $V(1)$, and $V_g(y_j)$. The
errors on Newton's data are displayed in 
Figure $2(b)$ while for CS binned data we take the $1-\sigma$ errors 
(which are half of the errorbars
displayed in Figure $2(b)$). For the gas we shall use  
a $20$ per cent error in the gas mass (see Section 4) which gives 
about 3 km s$^{-1}$ error for $V_g(y_j)$. The best-fit is obtained for

\begin{equation}
\beta_d(1)=0.42^{+0.15}_{-0.27}, \qquad
a=0.2^{+0.2}_{-0.2}, \qquad
c=1.3^{+0.3}_{-0.3}
\end{equation}
   
\begin{figure}
 
\vspace{10cm}
\includegraphics{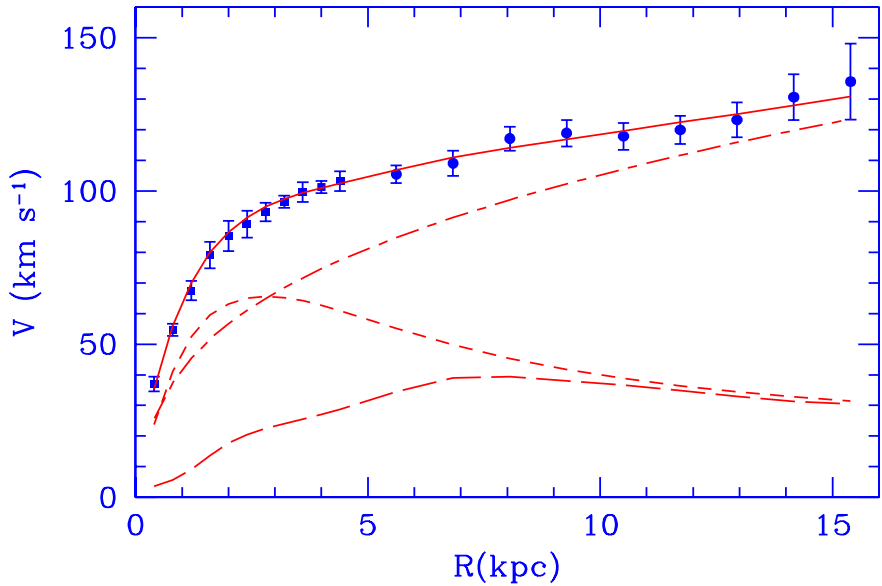}
\caption{ M33 rotation curve (points) compared with the best fit model
(continuous line). Also shown  the 
halo contribution (dashed-dotted line), the stellar disk (short dashed 
line) and  the gas contribution (long dashed line) }
\label{fig:fig6 }
\end{figure}

\noindent
The resulting rotational velocity, as well as the halo, gas,
and stellar contributions to it, is shown in Figure 6.
Similar values of the parameters are obtained if we perform the minimization
using the unbinned data or if we use the upper or lower limit
to $\beta_g(1)$. The best model reproduces, within $1-\sigma$ errorbars, 
all the velocity data; this is also evident from the low $\chi^2$ value: 
$\chi^2 = 9.0$ as defined in equation (6).
The quoted errors on the best fitting parameters correspond to the 
68 per cent confidence areas shown in black in Fig. 7. 
The best-fit model features the stellar disk as a major component 
of the gravitational potential for $R<2R_d$ with a dark halo becoming 
dominant at larger radii. The corresponding value of $M_d/L_B$ is
$0.8\pm 0.2$,  i.e. $M_d\simeq (3.4 \pm 0.8)\times 10^9$ M$_\odot$ 
($5$ per cent of which might be molecular gas).

\begin{figure}
 
\vspace{7cm}
\includegraphics{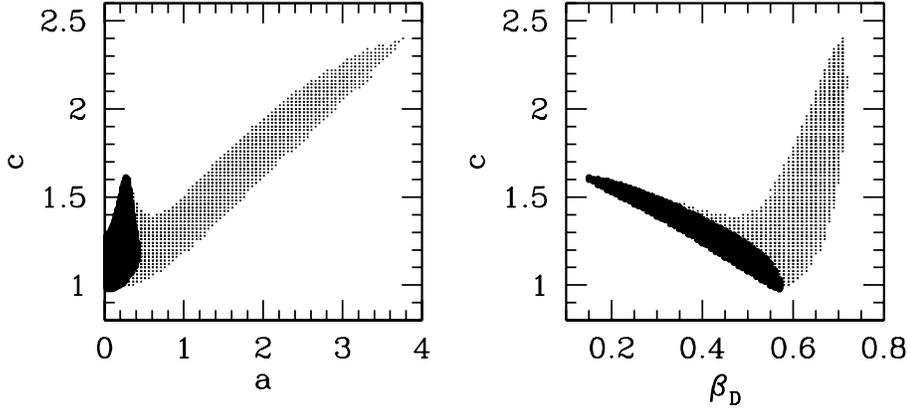}
\caption{ The $> 68\%$ confidence area in $a$-$c$, and $\beta_d$-$c$ plane.
Black areas are for halo model fit using all data points displayed in
Figure 6 while the black + light shaded
areas apply only if we exclude the innermost point from the fit.}
\label{fig:fig7 }
\end{figure}

\begin{figure}
 
\vspace{7cm}
\includegraphics{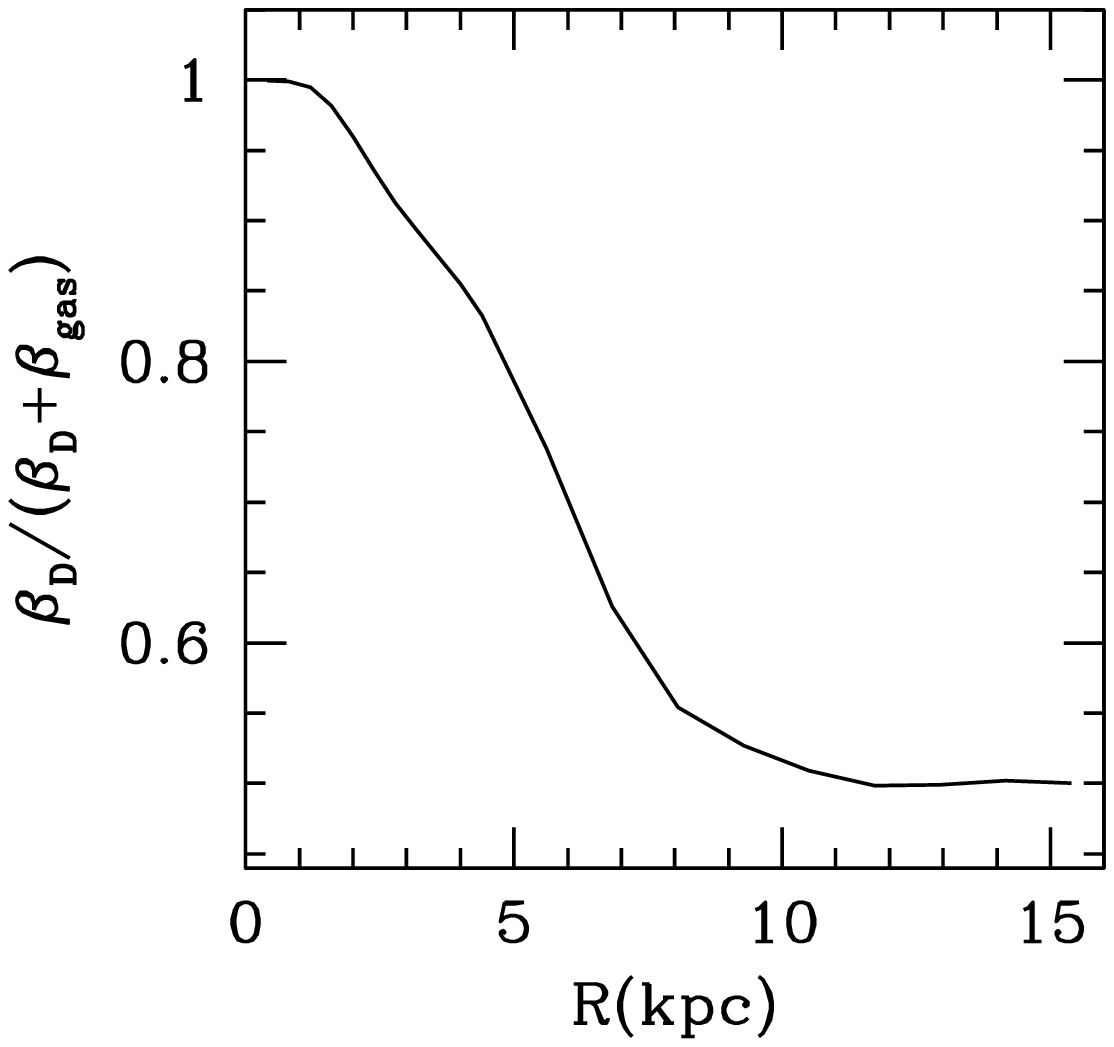}
\includegraphics{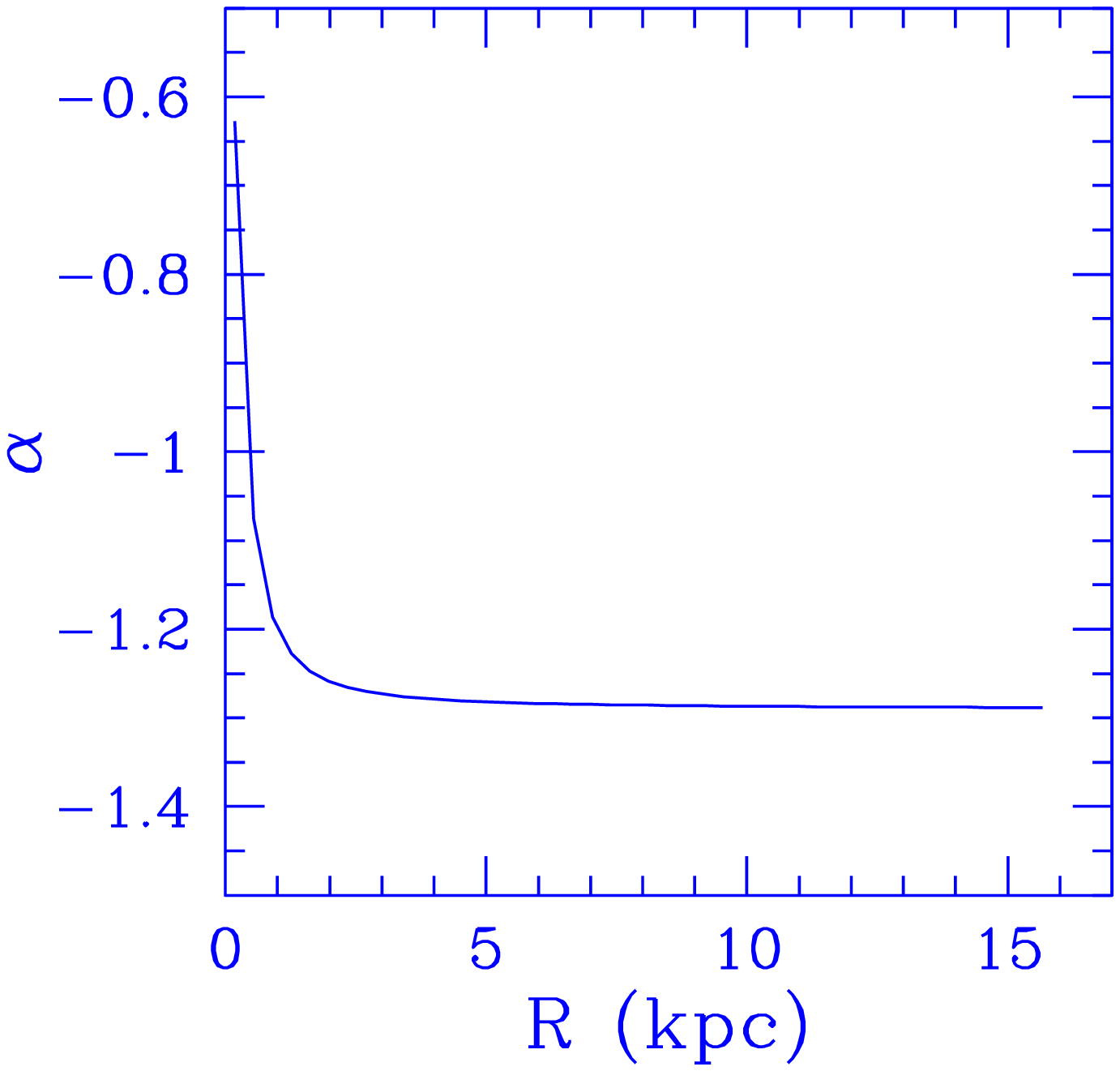}
\caption{The left hand panel shows the fractional contribution of the 
stellar disk velocity to the total velocity due to the all the luminous 
matter as a function of galactocentric radius (as from the best fit model).
The right hand panel shows the effective index of the dark matter density
radial decline.}
\label{fig:fig8 }
\end{figure}
 
The left hand panel of Figure 8 shows for the best-fit model the remarkable 
similarity between 
the \HI gas and the stellar contribution to $V^2$ beyond 8 kpc. 
The right hand panel of Figure 8 shows the effective radial decline of the 
dark matter density in the halo: $\rho_h(R)\propto R^{\alpha}$ 
with $\alpha\simeq -1.3$ throughout most of the galaxy. This agrees with 
the model-independent estimate given in Section 4, it proofs the validity of
the mass model, and excludes that the dark halo density declines as fast as
$R^{-2}$ in the outer \HI disk of M33.

\subsection
{Can we exclude the non-singular isothermal halo model?}

As a result of the regular rotation pattern, and of the small internal
errors, $\delta V/V \simeq 2-6$ per cent, the minimum $\chi^2$ fit   
puts aside disk-dominated models ($\beta_d(1)>0.6$) for M33 and
isothermal models for matter density of the dark halo  
($c\simeq 2$ ). Model-independent estimates of the dark matter in the
outermost regions of M33 confirm the shallower slope of the local density 
decline and its small related uncertainty. Here we would
like to examine additional sources of uncertainties which in principle
could make the non-singular isothermal halo model still compatible with
the data. First of all we would like to see if slight
variations of the inclination or position angle with 
respect to the best-fit tilted ring model can change the shape of the 
rotation curve sufficiently to favor a halo density with an effective 
decline $\alpha \simeq -2$
in regions where the outermost observable disk is. 
We therefore select variations of the deconvolution model which 
require $c\ge 1.3$ for the best-fit halo model. 
We show some of them in Figure 9 together with the best 
ring model (as in Figure 1). They all  give a  
symmetric rotation curve for the northern and southern part of M33  but
have higher values of the $\chi^2$ both 
for the ring model fit and for the best mass model fit.  
The corresponding values of $\beta_d$  and $c$ are shown as filled dots 
in Figure 9$(c)$  
(values of $a$ are all between 0.18 and 0.28). The open circles show 
$\beta_d(1)$  and $c$ obtained if we exclude
from the fit the last binned point of $V(R)$ (at 77 arcmin), which could in 
principle favor a rising slope of the rotation curve.
Noticeably, the points in Figure 9$(c)$ are all close 
or within the black confidence area shown in Figure 7. Therefore
within possible variations  of $i$ and $PA$ we can exclude for M33
a flat or declining rotation curve at the outermost observable 
galactocentric distances. 

\begin{figure}
 
\vspace{8.cm}
\includegraphics{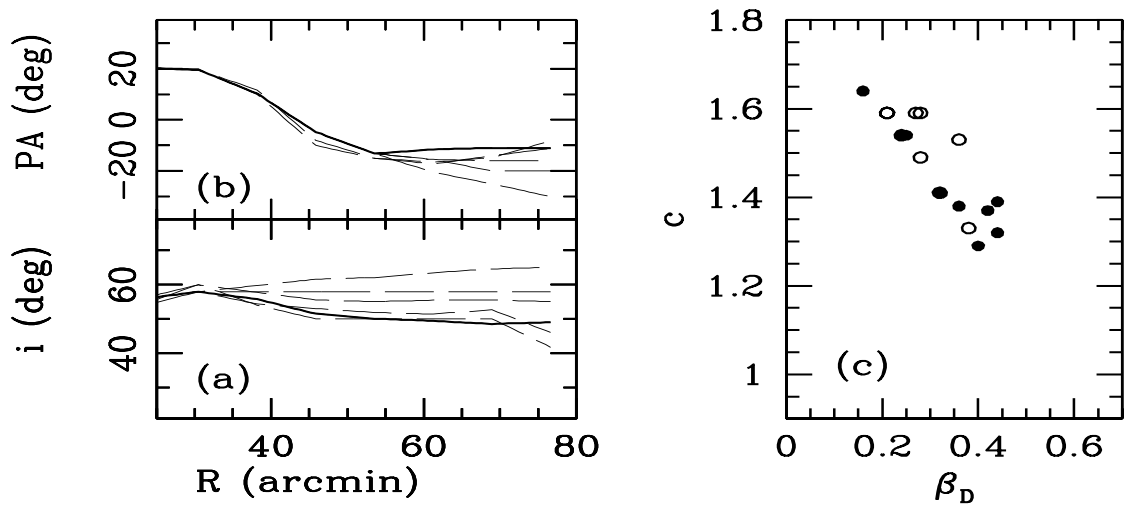}
\caption{Variations of the deconvolution parameters $i$ and $PA$
and the resulting values of $\beta_d$ and $c$ (filled dots). 
The open circles in $(c)$ refers to the same deconvolution parameters
but exclude the outermost binned point from the fit.}
\label{fig:fig9 }
\end{figure}

If the halo rotation curve is 
steadily and significantly increasing in the region 
where the dark matter is practically the unique mass component, this  
could be reconciled with the non-singular isothermal halo model only
if the M33 halo becomes fully isothermal well outside the observed
regions i.e. if the halo core, at constant density, extends for several kpc.
The results of the best-fit to the mass model presented in Section 5.1
seem to exclude this possibility. The correlation between $c$ and 
$\beta_d$ implies also that high
mass-to-light ratios for the stellar component correspond to high  
values of the dark matter velocity slopes. But what goes wrong if we 
now force $c=2$ ? In this case the best-fit values are $\beta_d(1)=
0.7$ and $a=2.6$. This mass model is able to reproduce the observed
rotation curve of M33 except close to the galaxy center where it predicts
a velocity much lower than value showed in Figure 2 for the innermost point. 
The low  spatial resolution of the velocity data used here prevents
the investigation of the rotation curve  for
$R< {1\over 2 }R_d$, where a more detailed model of
the molecular and stellar mass distribution are also needed. 
Wilson $\&$ Scoville (1989) have shown that, for $0.5\simlt R \simlt 1$ kpc,
the velocity field and the rotation curve derived
from the molecular hydrogen gas  are  similar
to those derived from the \HI observations at lower spatial resolution
and consistent with a rotational velocity proportional to $R^{0.5}$. 
For $R<0.3$ kpc however, velocities remain constant, at about 23 km s$^{-1}$,
consistently with optical data (Rubin $\&$ Ford 1985). As a consequence,
the value of the rotational velocity at 0.4 kpc results somewhat 
uncertain. If we minimize the $\chi^2$, defined by equation (6), excluding the 
innermost data point, we get the same values of  $a,\beta_d(1),c$ for
the best-fit model as in
Section 5.1, but quite larger uncertainties. The corresponding
$> 68$ per cent confidence areas are shown in Figure 7 as
light shaded areas. Our conclusion is that isothermal 
halo models with a flat rotation curve at $R\sim 4R_{opt}$ can be rejected
since the effective slope is in any case $\alpha \gg -2$; 
our spatial resolution is not sufficient to rule out isothermal
models with a large constant density core which would give 
a flat rotation curve only well beyond the observed outer \HI disk.

\subsection
{Testing the universal density profile in
hierarchical clustering models }

Recent calculations of cosmological structure formation in hierarchical 
clustering Cold Dark Matter models (e.g. Navarro, Frenk
$\&$ White 1996, 1997, Kravtsov, Klypin $\&$ Khokhlov 1997)
do not predict isothermal density distributions
but halos with a universal density profile which declines as $R^{-1}$ for 
$R<R_s$ and as $R^{-3}$
further out. $R_s$ is a scale radius which in terms of the virial radius
$R_{200}$ and of the ``concentration'' parameter $C$ can be written as
$R_s=R_{200}/C$. The resulting circular velocity $V$ has a well
defined analytic expression which depends on $C$ and on the circular 
velocity at the virial radius $V_{200}$ ($V_{200}$/km s$^{-1}$= $R_{200}/
h^{-1}$kpc where $h$ is the Hubble constant in units of 100 km s$^{-1}$
Mpc$^{-1}$). 
The rotation curve of M33 can be used to probe the validity of
universal density profile of the dark halo and its associated circular
velocity because it extends well beyond the optical radius, where the 
potential is more likely 
to be unchanged since the virialization epoch. Also,  the uncertainties
related to the mass of the luminous components in the inner regions
are small. Using $V_{200}$ and $C$ as free parameters we can find out if
the halo circular velocity in M33 can be well described by the density 
profile predicted by  hierarchical clustering Cold Dark Matter models.
In Figure 10 we show
the coincidence between the circular velocity of a halo formed 
in a hierarchical clustering Cold Dark Matter model for $V_{200}=127$ 
km s$^{-1}$, $C=8.4$, $h=0.5$, and the halo circular
velocity of the best-fit halo model for M33 discussed in Section 5.1.
Halo circular velocities for Cold Dark Matter models with
$0.5<h<1$ imply lower values of $C$ but they reproduce  
the best-fit halo model for M33 equally well (e.g. $V_{200}=137$ 
km s$^{-1}$, $C=5.6$ for $h=0.8$). The size
of the virialized halo is comparable to the distance between M33 and M31
suggesting that the two dark halos are touching. For other halo models 
lying in the $>68$ per cent contour areas of Figure
7, including the non-singular isothermal halo discussed in Section 5.2,
there are larger discrepancies with the profiles predicted by
Cold Dark Matter scenarios.

\begin{figure}
 
\vspace{8.cm}
\includegraphics{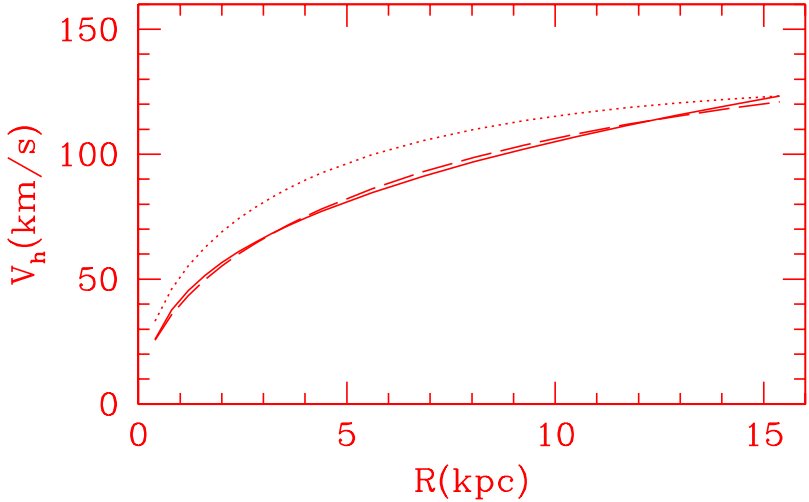}
\caption{Halo circular velocity of the best-fit mass model for M33 
presented Section 5.1 (continuous line) 
compared to the circular velocity of a halo formed in a hierarchical 
clustering Cold Dark Matter model for $V_{200}=127$ km s$^{-1}$, $C=8.4$, 
$h=0.5$ (dashed line).
The dotted line shows the halo velocity for the standard scenario
normalized to match the velocity in the outermost point. }
\label{fig:fig10 }
\end{figure}

The $C$ and $V_{200}$ values compatible with the M33 halo remain 
below the mass-concentration relation predicted for
the standard biased Cold Dark Matter scenario ($\Omega=1$, $\sigma_8=0.63$,
$h=0.5$, e.g. Navarro, Frenk $\&$ White 1997). The standard scenario in fact
requires $C=15$ to match the halo rotational velocity in the outermost point; 
the corresponding velocity profile is shown in Figure 10 where the
large discrepancies with the rotational velocities of the best-fit halo model 
are noticeable. 
The lower concentrations suggested by the M33 halo might indicate 
a low density universe with a non zero cosmological constant.  
A detailed discussion on the cosmological implications  
of our results requires however a more settled view of the 
the mass-concentration dependence on cosmological models and on   
environmental effects (Kravtsov, Klypin $\&$ Khokhlov 1997,
Navarro 1998, Avila-Rees, Firmani, Klypin $\&$ Kravtsov 1999).

\section{Conclusions}

The availability of \HI surface density data out to  large
distances from the center of the nearby galaxy M33 has allowed us
to derive the rotation curve for this galaxy out to $4\ R_{opt}$
and to  study in detail the dark matter distribution in the halo.
The results derived in the previous Sections lead to the 
following conclusions:
 
$\bullet$ As expected in a low-luminosity, late-type spiral, 
the \HI disk accounts for a substantial fraction of the total baryonic mass. 
It is remarkable that, the gas total mass   
($\sim 3\times 10^9$ M$_\odot$, including helium, molecular hydrogen,
atomic neutral and ionized hydrogen) is of the same order of the derived 
stellar mass for the best-fit mass model. This imply 
that $\sim 50$ per cent of the gas has turned into stars but the actual
mechanism that regulates the star formation activity in this blue galaxy
is still unclear since the \HI surface density is below the critical value
for a global disk instability. 

$\bullet$ The velocity data from 3 to 14 disk scale-lengths clearly  indicate 
that the rotation curve is steadily rising and is only weakly affected by the gravitational potential of the visible matter. The declining 
contribution of the combined visible baryonic components (gas+stars) to 
the observed velocity beyond $R\sim 8 $ kpc  excludes a direct correlation 
between the extended \HI surface density and the dark mass distribution.
M33 is therefore dark matter dominated and can
be used to directly probe the dark matter density at large 
galactocentric distances.

$\bullet$ The dark matter density, distributed in a spherical halo, 
decreases with radius as $R^{-1.3}$ with small uncertainties    
in the outermost observable regions, which more directly 
reflect the primordial dark matter distribution.  
This density profile agrees with the theoretical
predictions for halos formed in hierarchical clustering Cold Dark Matter 
models and implies a halo size comparable with the distance to M31. The 
halo is less centrally concentrated than expected from the
mass-concentration relation in the standard biased Cold Dark 
Matter scenario. 

\section*{Acknowledgments}

We would like to thank Leslie Hunt, Ed Salpeter, Renzo Sancisi,
Steve Schneider, and especially the anonymous referee for very
useful comments and a careful reading of the original manuscript.

\label{lastpage}

\end{document}